# Deep learning based cough detection camera using enhanced features

Gyeong-Tae Lee[a], Hyeonuk Nam[a], Seong-Hu Kim[a], Sang-Min Choi[a], Youngkey Kim[b], Yong-Hwa Park[a,*]

[a]Department of Mechanical Engineering, Korea Advanced Institute of Science and Technology, Daejeon 34141, South Korea

[b]SM Instruments Inc., Daejeon 34109, South Korea

* Corresponding author at: Department of Mechanical Engineering, Korea Advanced Institute of Science and Technology, Daejeon 34141, South Korea (Phone: +82-42-350-3235)

E-mail addresses: hansaram@kaist.ac.kr (G.T. Lee), frednam@kaist.ac.kr (H. Nam), seonghu.kim@kaist.ac.kr (S.H. Kim), cyanray1500@kaist.ac.kr (S.M. Choi), youngkey@smins.co.kr (Y. Kim), yhpark@kaist.ac.kr (Y.H. Park)

**ABSTRACT**

Coughing is a typical symptom of COVID-19. To detect and localize coughing sounds remotely, a convolutional neural network (CNN) based deep learning model was developed in this work and integrated with a sound camera for the visualization of the cough sounds. The cough detection model is a binary classifier of which the input is a two second acoustic feature and the output is one of two inferences (*Cough* or *Others*). Data augmentation was performed on the collected audio files to alleviate class imbalance and reflect various background noises in practical environments. For effective featuring of the cough sound, conventional features such as spectrograms, mel-scaled spectrograms, and mel-frequency cepstral coefficients (MFCC) were reinforced by utilizing their velocity (V) and acceleration (A) maps in this work. VGGNet, GoogLeNet, and ResNet were simplified to binary classifiers, and were named V-net, G-net, and R-net, respectively. To find the best combination of features and networks, training was performed for a total of 39 cases and the performance was confirmed using the test F1 score. Finally, a test F1 score of 91.9% (test accuracy of 97.2%) was achieved from G-net with the MFCC-V-A feature (named *Spectroflow*), an acoustic feature effective for use in cough detection. The trained cough detection model was integrated with a sound camera (i.e., one that visualizes sound sources using a beamforming microphone array). In a pilot test, the cough detection camera detected coughing sounds with an F1 score of 90.0% (accuracy of 96.0%), and the cough location in the camera image was tracked in real time.

Keywords: Cough detection, Coronavirus, COVID-19, Deep learning, Feature engineering, Sound visualization

## 1. Introduction

Since the COVID-19 outbreak started, there has been increasing demand for a monitoring system to detect human infection symptoms in real time in the field. The most common symptoms of infectious diseases including COVID-19 are fever and cough. While fever can be detected remotely using a thermal imaging camera, there is still no widespread monitoring system able to detect coughing. Since coughing is a major cause of virus transmission through airborne-droplets, it is very important to detect coughing to prevent the spread of infectious diseases. Although the cough detection is not sufficient to detect COVID-19, it is expected to be effective in preventing the spread of COVID-19 infection in the pandemic situation.

In previous studies to detect cough sounds, various acoustic features were used in conventional machine learning methods. Barry et al. (2006) developed a program that calculates characteristic spectral coefficients from audio recordings, which are then classified into cough and non-cough events by using probabilistic neural networks (PNN). Liu et al. (2013) introduced gammatone cepstral coefficients (GTCC) as a new feature and applied support vector machine (SVM) as a classifier for cough recognition. You et al. (2017a) extracted subband features by using gammatone filterbank and then trained SVM, k-nearest neighbors (k-NN) and random forest (RF) with the features in order to make final decision using ensemble method. Further, You et al. (2017b) exploited non-negative matrix factorization (NMF) to find the difference of cough and other sounds in a compact representation. In work by Drugman et al. (2013), signals including audio, thermistor, accelerometer, and electrocardiogram (ECG) were examined to find the best signal for cough detection, which turned out that audio signals are the most effective among them. In recent years, a variety of acoustic features have been evaluated and compared for use in cough detection (Miranda et al., 2019; Pramono et al., 2019). As deep neural network (DNN) has developed and shown excellent performance in fields such as vision, it became possible to detect cough sounds with high accuracy by applying DNN to acoustic features (Barata et al., 2019, 2020; Hossain et al., 2020). Recently, Ahamad et al. (2020) proposed a supervised model to identify early stage symptoms predicting COVID-19 disease, which revealed that fever (41.1%) and cough (30.3%) are the most frequent and significant predictive symptoms of COVID-19. Moreover, Laguarta et al. (2020) developed a convolutional neural network (CNN) based COVID-19 discriminator that only uses cough recordings as input. It is important to detect coughs accurately, but it is also necessary to observe the type and frequency components of coughs in real time to determine a patient's condition. Cough monitoring methods that quantify the number of coughs per unit of time have been proposed in several studies (Birring et al., 2008; Pramono et al., 2019; Wilhelm et al., 2003), and they are expected to show good monitoring performance when used together with previous cough detection algorithms.

However, since previous methods use cough sounds recorded in a limited place, it is vulnerable to general background noise when applied in practice, and impact sounds could be mistaken for cough sounds. In addition, since existing studies usually target one person, it is difficult to track the location of cough sounds in public places with many people. Therefore, in this study, we intend to propose a method that can detect cough sounds in a general environment and track it in real time through visualization of sound sources. To be applied in a real environment, sufficient learning about coughing sounds and various sound events in a general noisy environment should be carried out. Since coughing sound is a kind of transient sound event (Shin et al., 2009), it is necessary to construct an acoustic feature that is distinct from other sounds by highlighting the change in frequency components in a short time. In addition, since cough sound is clearly distinguished from other sounds with frequency change patterns and does not require a long-time context, it is effective to detect cough sounds based on CNN architecture. Lastly, in order to track coughing sounds in crowded public places, it is necessary to visualize their sound sources in real time. To detect human cough sounds in real time in the field, a CNN-based cough detection model was developed and integrated with a sound camera for visualization. The cough detection model determines whether a cough has occurred from the input audio signal, and the sound camera uses a beamforming technique to track the location of the cough and then to output an image showing the location with labeled contour on the camera screen. Although the proposed cough detection camera cannot detect COVID-19 directly, it can help disinfection against infectious diseases including COVID-19 by detecting and tracking cough sounds in real time in the field.

In this paper, the entire development process of the proposed deep learning-based cough detection camera is described. Section 2 shows the augmentation process of the dataset to be trained, and then describes the process of selecting the best model after training of candidate cough detection models, modified CNN architectures trained with various enhanced features. Section 3 explains the principle of a sound camera, and then describes the process of our cough detection camera after integration with the selected cough detection model. In Section 4, we investigated the feasibility of the proposed cough detection camera by comparing its performance with our previous prototype (Lee et al., 2020) through a pilot test, and the results of the pilot test are analyzed and discussed. Section 5 draws conclusion from previous discussion and outlines future applications.

## 2. Cough detection model

In this section, development procedure of the cough detection model is described. First, various sound event datasets were collected and data augmentation (DA) was applied to construct dataset for training, validation, and testing. Second, to detect cough patterns in detail, acoustic features were extracted from the dataset by combining spectrogram-based features and their derivatives. Third, three representative CNN architectures were simplified into binary classifiers and used as candidates for the cough detection model. Finally, a total of 39 cases were made by pairing each network and acoustic feature, and training was conducted on all candidate feature-network sets. The cough detection model was determined by selecting the best case (with the highest F1 score) on the test dataset. For reference, all development was carried out on Python, with the libraries of PyTorch, Torchaudio, and TQDM.

2.1. Dataset preparation

To train and evaluate the cough detection model, we collected seven datasets including HUMAN20200923, SMI-office, Google AudioSet (Gemmeke et al., 2017), NIGENS (Trowitzsch et al., 2020), DCASE2016-2 (Mesaros et al., 2018), DCASE2019-2 (Fonseca et al., 2020), and DCASE2018-5 (Dekkers et al., 2017). In Table 1, Datasets 1 and 2 are recordings of various sound events, including coughing sounds that could occur at home and office, respectively. Dataset 3 is a large-scale dataset of manually annotated audio events including coughing, sneezing, speech, and other sounds of human activities. Dataset 4 was provided for sound-related modeling in the field of computational auditory scene analysis (CASA), particularly for sound event detection. Dataset 5 is a training dataset available on the website of Task 2 of the 2016 IEEE AASP challenge on detection and classification of acoustic scenes and events (DCASE 2016). Dataset 6 is a collection of impact sounds from the Task 2 dataset of the DCASE 2019 challenge. A previous cough recognition model (Lee et al., 2020) had the problem of false positive errors, which in detail implies that other impulsive sounds were incorrectly recognized as cough sound. Thus, to supplement impulsive sound data in the training dataset, Dataset 6 was constructed by

Table 1

Collection of sound event datasets used for training and testing the cough detection model.

| No. | Dataset name | Class | Sampling rate (kHz) | Total (wav files) | Length (seconds) |
| --- | --- | --- | --- | --- | --- |
| 1 | HUMAN20200923 | Cough, dish, doorbell, general, glass, knock, phone, snoring, toilet. | 48.0 | 246 | 1,580 |
| 2 | SMI-office | Cough, speech, etc. | 44.1 | 1 | 439 |
| 3 | Google AudioSet | Cough, sneeze, sniffle, throat-clearing, speech, etc. | 44.1 | 1,171 | 50,920 |
| 4 | NIGENS | Alarm, baby, crash, dog, engine, female scream, female speech, fire, footsteps, general, knock, male scream, male speech, phone, piano. | 44.1 | 898 | 16,759 |
| 5 | DCASE2016-2 | Clearing throat, cough, door slam, drawer, keyboard, keys, knock, laughter, page turn, phone, speech. | 44.1 | 220 | 265 |
| 6 | DCASE2019-2 | Applause, cheering, chink & clink, clapping, crowd, dishes & pots & pans, fart, finger snapping, knock, shatter, slam, sneeze. | 44.1 | 888 | 4,336 |
| 7 | DCASE2018-5 | Absence, cooking, dishwashing, eating, other, social activity, vacuum cleaning, watching TV, working. | 16.0 | 72,984 | 729,837 |

selecting impact sounds such as clapping, knocking, and slamming from the DCASE 2019 Task 2 dataset. Dataset 7 is a development dataset of Task 5 of the DCASE 2018 challenge. It is suitable for background noise dataset because it is a recording of everyday activities performed at home (e.g., "watching TV").

To construct the dataset for the proposed model, Dataset 1 was used as the test dataset, datasets 2–6 as development datasets, and Dataset 7 as a background noise dataset for DA. Except for Dataset 7, all the wav files in datasets 1–6 were resampled to 16 kHz and divided into two classes, *Cough* and *Others*. Subsequently, they were cut at two second intervals. In this process, *Others* wav files were framed with 0% overlap, whereas *Cough* wav files had 50% overlap to obtain more cough data. As a result, the number of *Cough* and *Others* files was 145 and 683 (1:4.7) for the test dataset, and 5,952 and 32,144 (1:5.4) for the development dataset, respectively. In addition, DA was applied to mitigate the class imbalances in the development dataset and to maximize the robustness of the model to be used in real-world scenarios. To be specific, each *Cough* data was mixed with data from 45 different background noises, whereas each of the *Others* datasets was mixed with data from nine background noises. In this DA process, each piece of background noise data was randomly selected from the nine classes of Dataset 7, and data with zero energy were excluded. The mixed data of background noise data and sound event (*Cough* and *Others*) data were generated as follows:

$$\hat{x}(t) = \mu \cdot b(t)/b_{RMS} + (1 - \mu) \cdot s(t)/s_{RMS} \quad (1)$$

where $t$ denotes time index, $b(t)$ and $b_{RMS}$ are the waveform and root mean square (RMS) value of background noise data, respectively, $s(t)$ and $s_{RMS}$ are the waveform and RMS value of sound event data, respectively, and $\mu$ is a mixing ratio randomly selected between 0 and 0.4, which corresponds to the signal-to-noise ratio (SNR) from 3.5 dB to infinite (zero noise). To simulate different distances, the range of the mixed data was scaled as follows:

$$x(t) = \nu \cdot \hat{x}(t)/\max\{|\hat{x}(t)|\} \quad (2)$$

where $\max\{\}$ denotes the max operator, $||$ denotes the absolute operator, and $\nu$ is a volume ratio randomly selected between 0.6 and 1.0, where a volume ratio close to 0.6 means a longer distance. Finally, the training and validation datasets were organized by randomly dividing the augmented development dataset into two subsets at a 9 to 1 ratio. The number of wav files in the completed datasets to be used in the proposed model is shown in Table 2.

Table 2

Total number of wav files in the training, validation, and test datasets for the cough detection model.

|  | Train | Valid | Test | Total |
| --- | --- | --- | --- | --- |
| Cough | 240,858 | 26,762 | 145 | 267,765 |
| Others | 258,651 | 28,738 | 683 | 288,072 |
| Total | 499,509 | 55,500 | 828 | 555,837 |

2.2. Feature extraction

As mentioned above, since cough sound is a transient sound whose frequency characteristics change in a short time, it is necessary to build an acoustic feature in an image format suitable for pattern recognition of transient sounds. To effectively extract acoustic patterns from impulse-like cough sound, new acoustic features were designed using spectrograms (SP), mel-scaled spectrograms (MS), mel-frequency cepstral coefficients (MFCC), and their time derivatives, such as velocity map (V-map) and acceleration map (A-map).

SP is useful for time series pattern recognition because it represents changes in the frequency components of a signal with time. MS is another version of SP with a mel-scaled frequency range. The mel scale reflects the fact that humans do not perceive frequencies in a linear scale, and detect differences in lower frequencies better than in higher frequencies. Moreover, the mel-scaled frequency range approximates the response of the human auditory system more closely than linearly spaced frequency range does. Notably, the gaps between the harmonic components in MS are equally spaced even if the pitch moves. An MFCC is a mel-scaled version of the cepstrum, and a feature widely used in automatic speech and speaker recognition. The sounds from the human mouth are filtered by the vocal tract, including tongue and teeth. If their response could be determined accurately, this would give an accurate representation of what sounds are being produced. The shape of the response appears on the envelope of the spectrum, and the role of MFCC is to accurately represent this envelope. Because a coughing sound is essentially a transient sound, it is effective to reflect the rate of change over time in acoustic feature to distinguish it from the other similar transient sounds, such as cough-like sounds or impulsive sounds. Therefore, a proposal was made to use V-map and A-map, which can be derived from spectrogram-based features such as SP, MS, and MFCC. The V-map represents the amount of change in a spectrogram-based feature over time, and is composed of forward difference (at $t = 0$), central difference (at $1 \leq t \leq T - 1$), and backward difference (at $t = T$) as follows:

$$V(t,f) = \begin{cases} X(t+1,f) - X(t,f); & (t = 0), \\ \dfrac{X(t+1,f) - X(t-1,f)}{2}; & (1 \leq t \leq T - 1), \\ X(t,f) - X(t-1,f); & (t = T), \end{cases} \quad (3)$$

where $f$ the frequency (or coefficient) index, $X(t,f)$ the spectrogram-based feature, and $T$ the time length of the feature. The A-map represents the amount of change in the V-map over time, and is defined similarly to V-map as

$$A(t,f) = \begin{cases} V(t+1,f) - V(t,f); & (t = 0), \\ \dfrac{V(t+1,f) - V(t-1,f)}{2}; & (1 \leq t \leq T - 1), \\ V(t,f) - V(t-1,f); & (t = T). \end{cases} \quad (4)$$

For the invariance of feature to loudness, the waveform of input data was normalized to be within −1.0 to +1.0 by maximum value division. Then, the normalized waveform was converted to SP, MS, and MFCC with an FFT bin size of 512, and with window and hop sizes of 30 ms and 15 ms, respectively. Because coughing sounds have a spectral distribution between 350 Hz and 4 kHz (Shin et al., 2009), the maximum frequency of these features was set to at least 4 kHz. In the end, SP, MS, and MFCC were concatenated with their respective V-maps and A-maps to create a variety of features with dimensions of 3 (channel) × 128 (height) × 128 (width).

As an example, the waveform of a coughing sound is shown in Fig. 1(a). Fig. 1(b) is an SP-based three-channel acoustic feature (SP-V-A) that concatenates the SP with its V-map and A-map. Similarly, Fig. 1(c) is an MS-based three-channel acoustic feature (MS-V-A), and Fig. 1(d) is an MFCC-based three-channel acoustic feature (MFCC-V-A). For comparison with a coughing sound, acoustic features of cough-like sounds, such as sneezing and throat-clearing, are shown in Figs. 2 and 3, and those of impulsive sounds, such as clapping and knocks, are shown in Figs. 4 and 5. In Fig. 1 (cough), a similar spectral pattern is repeated two or three times, and distinct vertical lines appear at the beginning and end of the cough over the entire frequency range. In addition, tonal components related to the vocal tract stand out between the vertical lines. In Fig. 2 (sneeze), the feature's magnitude is small when breathing in the air before 0.75 seconds, whereas the magnitude becomes very large when ejecting droplets after one second. In addition, vertical lines appear clearly at the beginning of droplet ejection. In Fig. 3 (throat-clearing), the feature is mainly composed of an initial short and later a long pattern. It is noticeable that the tonal components of the vocal tract shift to lower frequencies later. An abrupt vertical line is evident at the start of the event. In Fig. 4 (clapping), the feature consists of an overlapped series of typical impulse responses. Each distinct vertical line appears over the entire frequency range at the start of each clap. In Fig. 5 (knock), the feature consists of several impulse responses, and energy is concentrated mainly at low frequencies. An abrupt vertical line occurs only at the beginning of each event. As described above, it could be confirmed with the naked eye that the difference between each sound event clearly appeared on these acoustic features. These features were then passed through the multilayer CNN models.

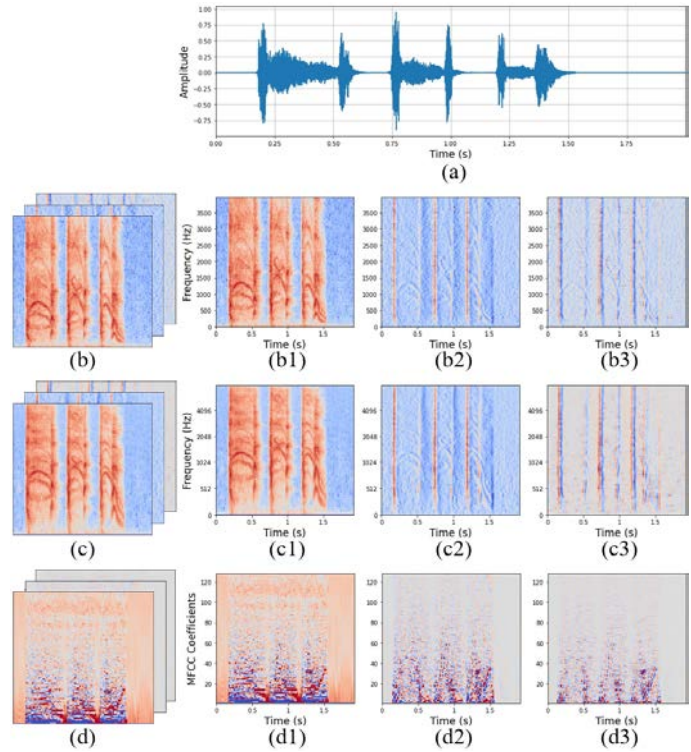

Fig. 1. Waveform and features of cough sound: (a) Waveform, (b) SP-V-A, (b1) SP, (b2) V-map of SP, (b3) A-map of SP, (c) MS-V-A, (c1) MS, (c2) V-map of MS, (c3) A-map of MS, (d) MFCC-V-A, (d1) MFCC, (d2) V-map of the MFCC, (d3) A-map of the MFCC.

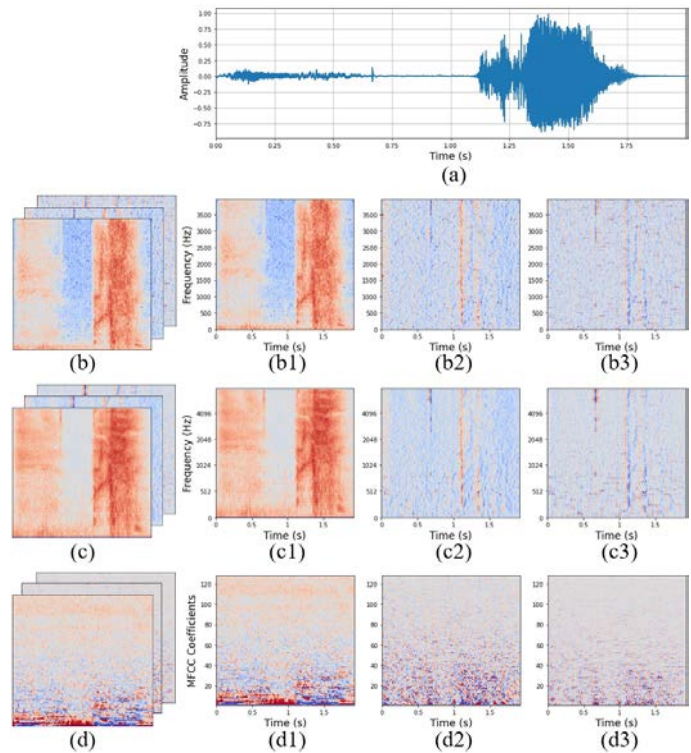

Fig. 2. Waveform and features of a sneeze sound: (a) Waveform, (b) SP-V-A, (b1) SP, (b2) V-map of SP, (b3) A-map of SP, (c) MS-V-A, (c1) MS, (c2) V-map of MS, (c3) A-map of MS, (d) MFCC-V-A, (d1) MFCC, (d2) V-map of the MFCC, (d3) A-map of the MFCC.

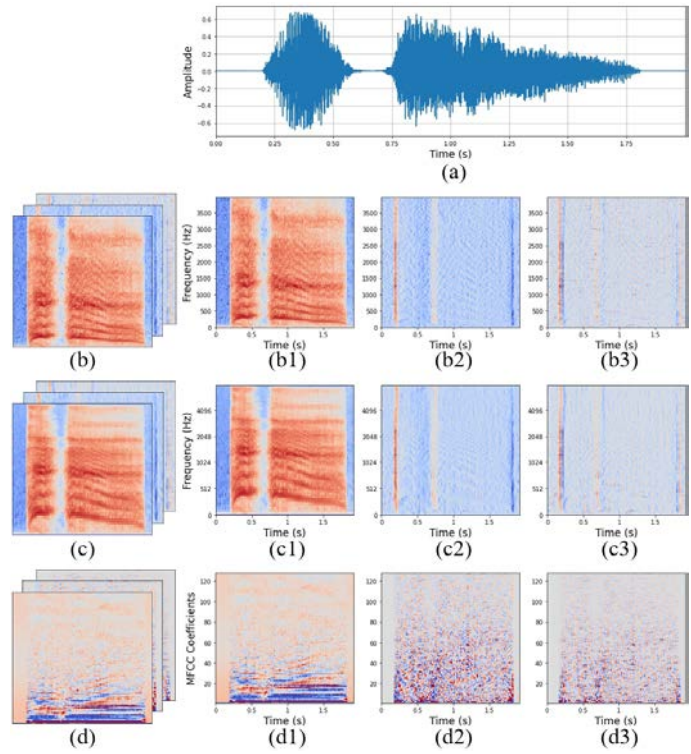

Fig. 3. Waveform and features of a throat-clearing sound: (a) Waveform, (b) SP-V-A, (b1) SP, (b2) V-map of SP, (b3) A-map of SP, (c) MS-V-A, (c1) MS, (c2) V-map of MS, (c3) A-map of MS, (d) MFCC-V-A, (d1) MFCC, (d2) V-map of the MFCC, (d3) A-map of the MFCC.

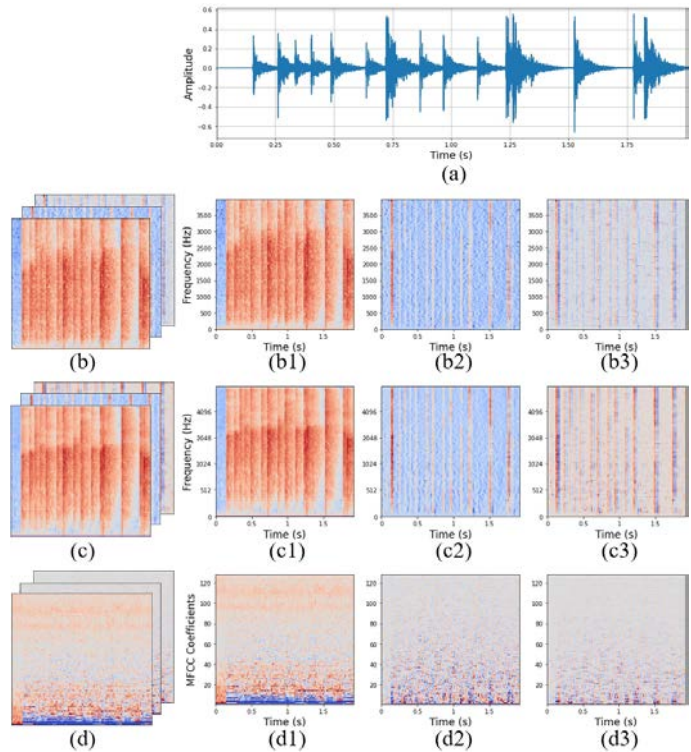

Fig. 4. Waveform and features of a clapping sound: (a) Waveform, (b) SP-V-A, (b1) SP, (b2) V-map of SP, (b3) A-map of SP, (c) MS-V-A, (c1) MS, (c2) V-map of MS, (c3) A-map of MS, (d) MFCC-V-A, (d1) MFCC, (d2) V-map of the MFCC, (d3) A-map of the MFCC.

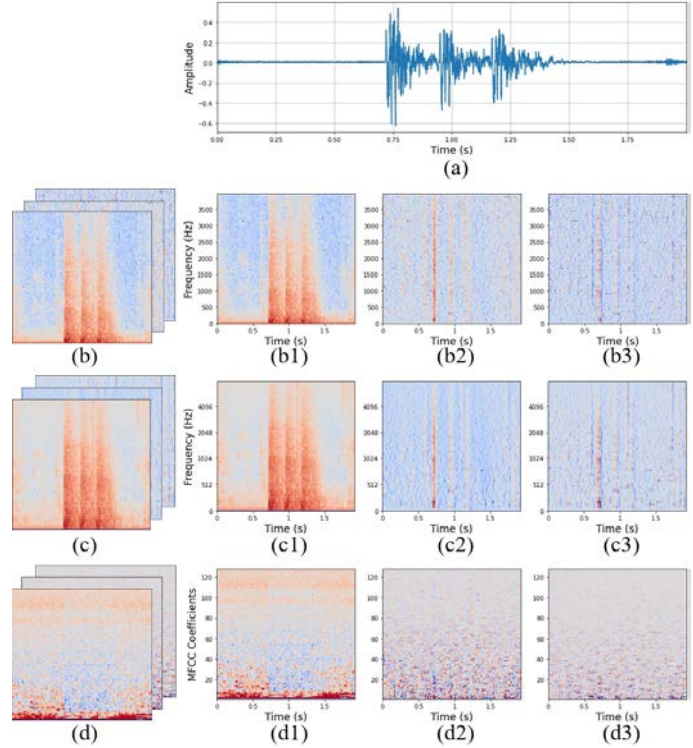

Fig. 5. Waveform and features of knocking sound: (a) Waveform, (b) SP-V-A, (b1) SP, (b2) V-map of SP, (b3) A-map of SP, (c) MS-V-A, (c1) MS, (c2) V-map of MS, (c3) A-map of MS, (d) MFCC-V-A, (d1) MFCC, (d2) V-map of the MFCC, (d3) A-map of the MFCC.

2.3. CNN architecture

The cough detection model is a binary classifier where the input is a two-second long acoustic feature and the output is a *Cough* or *Others* inference. For binary classification, simplified representative CNN models such as VGGNet (Simonyan & Zisserman, 2015), GoogLeNet (Szegedy et al., 2015), and ResNet (He et al., 2016), were renamed V-net, G-net, and R-net, respectively in this work.

2.3.1. V-net

VGGNet is the CNN model proposed by Simonyan and Zisserman from the Oxford's Visual Geometry Group (VGG). Although it ranked second after GoogLeNet in ILSVRC 2014, it is used more than GoogLeNet because it is structurally simple, easy to understand, and useful for transformation. VGGNet was originally designed to examine the effect of network depth on performance. Therefore, those workers fixed the kernel size to $3 \times 3$ and increased the number of convolutional (conv) layers to create six structures (A, A-LRN, B, C, D, E) and compared their performance. A $3 \times 3$ kernel can make the network deeper than a $5 \times 5$ or $7 \times 7$ kernel, and this has the effect of increasing the learning speed by reducing the number of parameters. In addition, as the number of conv layers increases, more ReLU activation functions are used. This increases nonlinearity, so more effective features can be extracted. In fact, when 3 × 3 convolution was used twice instead of using 5 × 5

Table 3

V-net configuration for binary classification.

| Type | Kernel size / Stride | Output size (C × H × W) |
|---|---|---|
| Conv+GroupNorm+ReLU | 3 × 3 / 1 | 16 × 128 × 128 |
| Conv+GroupNorm+ReLU | 3 × 3 / 1 | 16 × 128 × 128 |
| MaxPool | 2 × 2 / 2 | 16 × 64 × 64 |
| Conv+GroupNorm+ReLU | 3 × 3 / 1 | 32 × 64 × 64 |
| Conv+GroupNorm+ReLU | 3 × 3 / 1 | 32 × 64 × 64 |
| MaxPool | 2 × 2 / 2 | 32 × 32 × 32 |
| Conv+GroupNorm+ReLU | 3 × 3 / 1 | 64 × 32 × 32 |
| Conv+GroupNorm+ReLU | 3 × 3 / 1 | 64 × 32 × 32 |
| Conv+GroupNorm+ReLU | 3 × 3 / 1 | 64 × 32 × 32 |
| MaxPool | 2 × 2 / 2 | 64 × 16 × 16 |
| Conv+GroupNorm+ReLU | 3 × 3 / 1 | 128 × 16 × 16 |
| Conv+GroupNorm+ReLU | 3 × 3 / 1 | 128 × 16 × 16 |
| Conv+GroupNorm+ReLU | 3 × 3 / 1 | 128 × 16 × 16 |
| MaxPool | 2 × 2 / 2 | 128 × 8 × 8 |
| Conv+GroupNorm+ReLU | 3 × 3 / 1 | 128 × 8 × 8 |
| Conv+GroupNorm+ReLU | 3 × 3 / 1 | 128 × 8 × 8 |
| Conv+GroupNorm+ReLU | 3 × 3 / 1 | 128 × 8 × 8 |
| MaxPool | 2 × 2 / 2 | 128 × 4 × 4 |
| FullyConnected+ReLU |  | 512 × 1 × 1 |
| FullyConnected+ReLU |  | 32 × 1 × 1 |
| FullyConnected |  | 2 × 1 × 1 |
| Softmax |  | 2 × 1 × 1 |

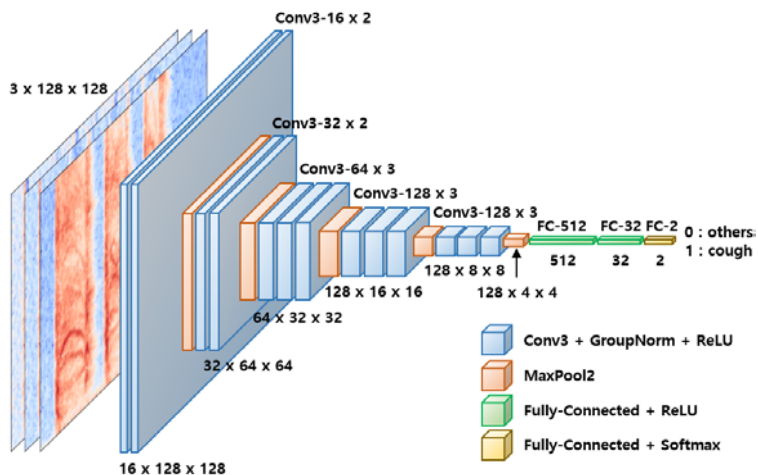

Fig. 6. Architecture of a cough detection model based on V-net.

convolution, the performance improved by about 7% (Simonyan & Zisserman, 2015). However, VGGNet is disadvantaged by having too many parameters. The main reason is the use of three fully connected (FC) layers in the final stage, so that the number of parameters in this part is about 122 million.

Therefore, the total number of neurons in the FC layers were reduced to about 1/17 to alleviate this shortcoming. In addition, the number of kernels in each conv layer was reduced to 1/4 based on a D-type VGGNet to be suitable for binary classification. Finally, a V-net was made by applying group normalization (GN) (Wu & He, 2018) after each conv layer so that the model performance would be robust against difference in the batch size. The configuration and architecture of V-net are outlined in Table 3 and Fig. 6, respectively.

2.3.2. G-net

Szegedy *et al.* from Google developed the Inception module to improve performance by deepening the CNN architecture without increasing the amount of computation. GoogLeNet, which won the ILSVRC 2014, is one of several versions of Inception-based architecture (Szegedy et al., 2015). GoogLeNet can be divided into four parts: pre-layer, Inception modules, average pooling, and auxiliary classifier. The pre-layer is a typical CNN part that exists for initial learning. Because the learning efficiency of the Inception module decreases near the input part, the pre-layer was added for learning convenience. After the pre-layer, a total of nine Inception modules follow. Each Inception module extracts feature maps from $1 \times 1$, $3 \times 3$, and $5 \times 5$ kernels; then, concatenates them in the channel direction, and delivers the result to the next layer. This differentiates this CNN from conventional one that uses the same kernel size in the same layer, and allows more diverse types of characteristics to be derived. However, in this case, the total amount of computation increases. To compensate for this, $1 \times 1$ conv layers are placed in front of the $3 \times 3$ and $5 \times 5$ kernels to reduce the number of input feature maps. As the number of feature maps decreases, the amount of computation decreases, so the network can be built deeper. In the classification process, applying average pooling instead of an FC layer after the final

Table 4

G-net configuration for binary classification.

| Type | Kernel size / Stride | Output size (C × H × W) | #1 × 1 | #3 × 3 reduce | #3 × 3 | #5 × 5 reduce | #5 × 5 | Pool proj |
|---|---|---|---|---|---|---|---|---|
| Conv+GroupNorm+ReLU | 7 × 7 / 1 | 16 × 128 × 128 | | | | | | |
| MaxPool | 3 × 3 / 2 | 16 × 64 × 64 | | | | | | |
| Conv+GroupNorm+ReLU | 3 × 3 / 1 | 48 × 64 × 64 | | | 48 | | | |
| MaxPool | 3 × 3 / 2 | 48 × 32 × 32 | | | | | | |
| Inception Module 1 | | 256 × 32 × 32 | 64 | 96 | 128 | 16 | 32 | 32 |
| Inception Module 2 | | 480 × 32 × 32 | 128 | 128 | 192 | 32 | 96 | 64 |
| MaxPool | 3 × 3 / 2 | 480 × 16 × 16 | | | | | | |
| Inception Module 3 | | 512 × 16 × 16 | 192 | 96 | 208 | 16 | 48 | 64 |
| Inception Module 4 | | 512 × 16 × 16 | 160 | 112 | 224 | 24 | 64 | 64 |
| Inception Module 5 | | 512 × 16 × 16 | 128 | 128 | 256 | 24 | 64 | 64 |
| Inception Module 6 | | 528 × 16 × 16 | 112 | 144 | 288 | 32 | 64 | 64 |
| Inception Module 7 | | 832 × 16 × 16 | 256 | 160 | 320 | 32 | 128 | 128 |
| MaxPool | 3 × 3 / 2 | 832 × 8 × 8 | | | | | | |
| Inception Module 8 | | 832 × 8 × 8 | 256 | 160 | 320 | 32 | 128 | 128 |
| Inception Module 9 | | 1024 × 8 × 8 | 384 | 192 | 384 | 48 | 128 | 128 |
| AvgPool | 8 × 8 / 1 | 1024 × 1 × 1 | | | | | | |
| FullyConnected | | 2 × 1 × 1 | | | | | | |
| Softmax | | 2 × 1 × 1 | | | | | | |

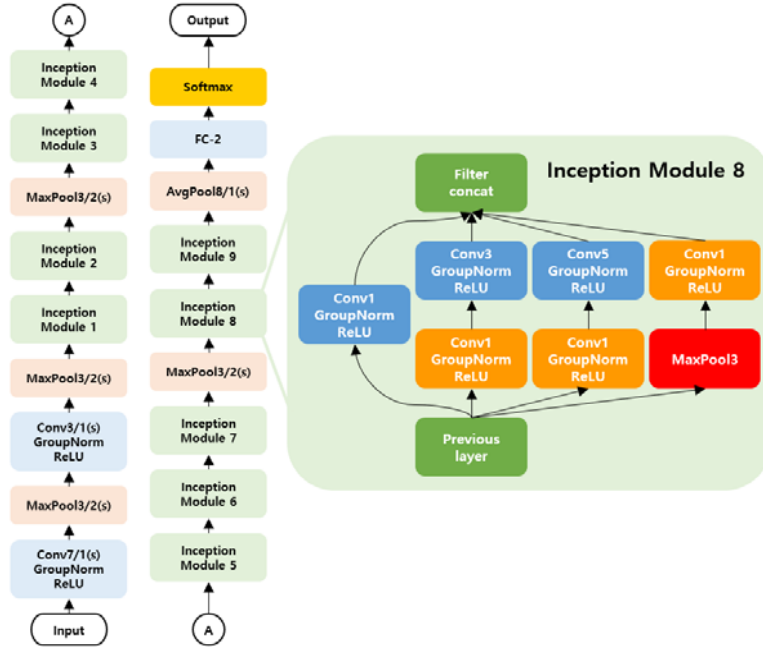

Fig. 7. Architecture of a cough detection model based on G-net.

Inception module has the effect of significantly reducing the computational amount. Average pooling does not require a learning process; therefore, no additional parameters are generated. Softmax is used as a classifier, and GoogLeNet has one at the end and two in the middle. The two classifiers in the middle (called auxiliary classifiers) alleviate the vanishing gradient problem and help make network convergence quick. Their losses during training are weighted (0.3) and are added to the total loss. After training, these middle classifiers are discarded for the test stage.

The GoogLeNet was modified as a binary classifier and named G-net. First, the pre-layer was changed to fit the size of the input feature, and the number of FC layer nodes was reduced to two for binary classification. In later preliminary experiments, it was found that the G-net does not have the vanishing gradient problem even without the auxiliary classifiers. Therefore, to exclude the effect on the total loss during training, the auxiliary classifiers were removed and only Softmax was used at the end. In addition, the dropout layer after the average pooling was removed. Dropout is an effective regularization method, but it is not efficient in terms of resource usage because computing devices are optimized for dense matrix operations. Finally, GN was added after each conv layer so that the model's performance would not be affected by batch size. The configuration and architecture of the G-net are outlined in Table 4 and Fig. 7, respectively.

2.3.3. R-net

He *et al.* from Microsoft found that, when the depth exceeds a certain level, the performance of the deep model becomes lower than that of the shallow model. Therefore, they proposed a method called residual learning as a way to solve this problem and won the ILSVRC 2015 with the ResNet that implemented it (He et al., 2016). The residual block, the core of ResNet, adds its output to its input transmitted through the skip connection and passes it to the next step. A residual block has the advantage of using both the simple characteristics of the input terminal and the complex characteristics of the output terminal. Moreover, the slope of the addition operation is "1" when performing backpropagation. Consequently, the vanishing gradient problem can be solved by effectively propagating the loss to the front part of the model. For ResNet with over 50 layers, a bottleneck block adopting the $1 \times 1$ convolution of the Inception module was used as a building block to reduce complexity and model size. The bottleneck block first compresses the channel of the input feature with $1 \times 1$ convolution. Then, a feature is extracted by $3 \times 3$ convolution, the channel is expanded again by 1 × 1 convolution, and the input feature delivered through skip connection is added and transferred to the next layer. For reference, the complexity is less and the model size smaller in the 152-layer ResNet than in VGGNet. Last, batch normalization (BN) was adopted after each conv layer and before the activation function.

Table 5

R-net configuration for binary classification.

| Type | Kernel size, C / Stride | Output size (C × H × W) |
| --- | --- | --- |
| Conv+GroupNorm+ReLU | 7 × 7, 16 / 1 | 16 × 128 × 128 |
| MaxPool | 3 × 3, 16 / 2 | 16 × 64 × 64 |
| Bottleneck Block Set 1 | $\begin{bmatrix} 1 \times 1, 16 \\ 3 \times 3, 16 \\ 1 \times 1, 64 \end{bmatrix} \times 3$ | 64 × 32 × 32 |
| Bottleneck Block Set 2 | $\begin{bmatrix} 1 \times 1, 32 \\ 3 \times 3, 32 \\ 1 \times 1, 128 \end{bmatrix} \times 4$ | 128 × 16 × 16 |
| Bottleneck Block Set 3 | $\begin{bmatrix} 1 \times 1, 64 \\ 3 \times 3, 64 \\ 1 \times 1, 256 \end{bmatrix} \times 6$ | 256 × 8 × 8 |
| Bottleneck Block Set 4 | $\begin{bmatrix} 1 \times 1, 128 \\ 3 \times 3, 128 \\ 1 \times 1, 512 \end{bmatrix} \times 3$ | 512 × 8 × 8 |
| AvgPool | 8 × 8, 512 / 1 | 512 × 1 × 1 |
| FullyConnected | | 2 × 1 × 1 |
| Softmax | | 2 × 1 × 1 |

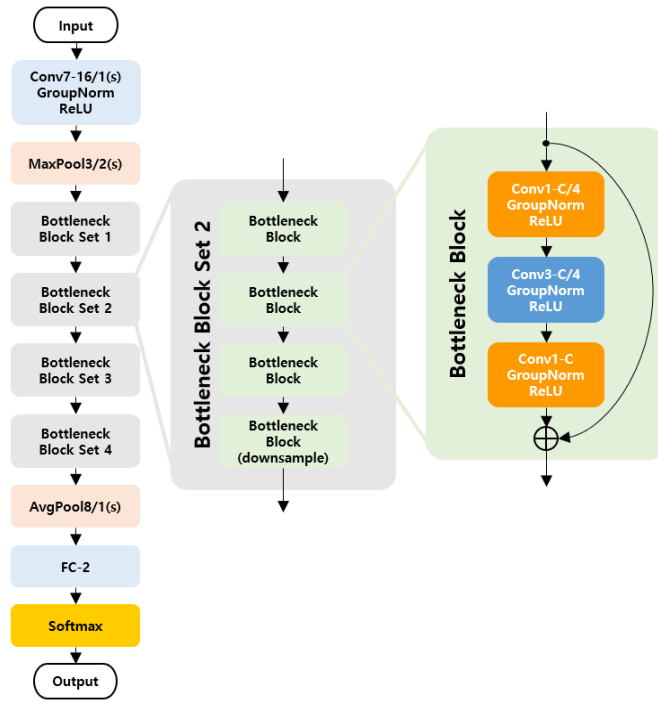

Fig. 8. Architecture of a cough detection model based on R-net.

Based on the 50-layer ResNet, the number of kernels in each conv layer was reduced to 1/4 for binary classification. In addition, the conv layer at the input part was modified to fit the size of the input feature, and the FC layer of the output part was modified (became two-way). Finally, the R-net was built by applying GN instead of BN; so the performance of the model was not affected by batch size. The configuration and architecture of the R-net are outlined in Table 5 and Fig. 8, respectively.

2.4. Modeling procedure

Various candidate groups were created for the cough detection model by combining the features and networks proposed above. In Table 6, Type 1 is a model candidate group in which V-net, G-net, and R-net were trained using the basic features SP, MS, and MFCC, respectively. Types 2 and 3 are model candidate groups trained with two-channel features applying V-map, and three-channel features applying both V-map and A-map, respectively. By comparing the results of Types 1 to 3, the effect of V-map and A-map on the performance of the models was examined. Moreover, Types 4 and 5 were organized to examine the effect of combinations of basic features on the performance of the models. Types 4 and 5 are model candidate groups trained with two-channel features and three-channel features, respectively. These groups reflect possible combinations of the basic features.

Table 6

List of candidate groups for cough detection model.

| Type | Model summary | Cases |
| --- | --- | --- |
| 1 | Feature: SP, MS, MFCC<br>Network: V-net, G-net, R-net | 9 |
| 2 | Feature: SP-V, MS-V, MFCC-V<br>Network: V-net, G-net, R-net | 9 |
| 3 | Feature: SP-V-A, MS-V-A, MFCC-V-A<br>Network: V-net, G-net, R-net | 9 |
| 4 | Feature: SP-MS, SP-MFCC, MS-MFCC<br>Network: V-net, G-net, R-net | 9 |
| 5 | Feature: SP-MS-MFCC<br>Network: V-net, G-net, R-net | 3 |

Accuracy, recall, precision, and F1 score were used as the performance metrics of each model. These were defined as

$$Accuracy = \frac{TP + TN}{TP + TN + FP + FN} \quad (5)$$

$$Recall = \frac{TP}{TP + FN} \quad (6)$$

$$Precision = \frac{TP}{TP + FP} \quad (7)$$

$$F_1 = 2 \cdot \frac{Precision \cdot Recall}{Precision + Recall} \quad (8)$$

where $TP$ denotes true positive, $TN$ true negative, $FP$ false positive, and $FN$ false negative. In this case, positive and negative refer to *Cough* and *Others*, respectively. Because accuracy can be a misleading metric for an imbalanced dataset, the final model with the highest F1 score was selected as the test dataset among 39 candidate models of Types 1 to 5, and confirmed it as the cough detection model.

The deep learning server was equipped with $1 \times$ Intel® i9-10980XE CPU, $4 \times$ NVIDIA® TITAN RTX GPUs, and 256 GB RAM. The batch size and the maximum number of epochs were set to 256 and 100, respectively. To ensure convergence, all input feature batches were normalized per channel using the mean and standard deviation per channel of the training dataset features before entering each network. Each candidate model was trained by minimizing the cross-entropy loss using the Adam optimizer with a learning rate of $1 \times 10^{-3}$. The learning rate scheduler was set to reduce the learning rate by a factor of 0.1 when the validation loss did not decrease for five epochs. In addition, to avoid overfitting, the training was set to stop when the learning rate fell below $1 \times 10^{-5}$.

The performances of Types 1 to 5 on the test dataset after training are shown in Tables 7 to 11. In each table, the top three F1 scores are in bold font, and only the top one F1 score is underlined.

In Table 7, the top three Type 1 models are MS & G-net (85.2%), SP & G-net (84.0%), and MS & R-net (83.0%). All of the top three models have higher precision than recall. In particular, MS & G-net, the top model, has the highest precision among the top three, while having the lowest recall. Fig. 9 shows the training history of MS & G-net and its normalized confusion matrix for the test dataset. The training converged from the 31st epoch and ended at the 43rd epoch. In Fig. 9(f), normalized $FN$ (0.23) is considerably higher than normalized $FP$ (0.01), which means that *Cough* is often mistaken for *Others*. Thus, Type 1 is not suitable as a cough detection model.

In Table 8, the top three Type 2 models are MS-V & G-net (90.0%), SP-V & G-net (86.5%), and MS-V & R-net (85.5%). All of the Type 2 models have higher precision than recall, and the best model, MS-V & G-net, has the highest precision. Fig. 10 shows the training history of MS-V & G-net and its normalized confusion matrix for the test dataset. The training converged from the 23rd epoch and ended at the 35th epoch. As shown in Fig. 10(f), normalized $FN$ (0.17) was less than MS & G-net of Type 1, and normalized $FP$ became 0.00. This means that *Others* is almost classified as *Others*. Therefore, MS-V & G-net is better than MS & G-net.

In Table 9, the top three Type 3 models are MFCC-V-A & G-net (91.9%), MS-V-A & G-net (90.2%), and MS-V-A & R-net (88.7%). All of the top three models have higher precision than recall, and the top model, MFCC-V-A & G-net, is the only model that exceeds 90.0% in both recall and precision. Fig. 11 shows the training history of MFCC-V-A & G-net and its normalized confusion matrix for the test dataset. The training converged from the 27th epoch and ended at the 39th epoch. In Fig. 11(f), normalized $FP$ (0.01) slightly increased compared to Type 2, but normalized TP became 0.90, indicating that *Cough* is better classified than with Type 2. Accordingly, MFCC-V-A & G-net is more suitable as a cough detection model than MS-V & G-net is.

In Table 10, the top three Type 4 models are SP-MS & G-net (89.9%), SP-MS & R-net (88.1%), and SP-MFCC & G-net (87.8%). The best model, SP-MS & G-net, has the highest recall in Type 4, and the other models have higher precision than recall. Fig. 12 shows the training history of SP-MS & G-net and its normalized confusion matrix for the test dataset. The training converged from the 35th epoch and ended at the 46th epoch. Compared with MS-V & G-net of Type 2, Fig. 12(f) shows that normalized $FN$ (0.08) is lower, but normalized $FP$ (0.03) is higher. Consequently, SP-MS & G-net cannot be considered better than MS-V & G-net.

In Table 11, the top model of Type 5 is SP-MS-MFCC & R-net (90.1%). The top model has the smallest gap between recall and precision in Type 5. Fig. 13 shows the training history of SP-MS-MFCC & R-net and its normalized confusion matrix for the test dataset. The training converged from the 32nd epoch and ended at the 42nd epoch. In Fig. 13(f), compared to SP-MS & G-net of Type 4, normalized $FP$ (0.02) slightly decreased,

Table 7

Performance of Type 1 models in terms of Accuracy (A), Recall (R), Precision (P), and F1 score (F1) on the test dataset.

| Models | SP | | | | MS | | | | MFCC | | | |
|---|---|---|---|---|---|---|---|---|---|---|---|---|
| | A (%) | R (%) | P (%) | F1 (%) | A (%) | R (%) | P (%) | F1 (%) | A (%) | R (%) | P (%) | F1 (%) |
| V-net | 93.1 | 84.1 | 78.2 | 81.1 | 94.2 | 70.3 | 95.3 | 81.0 | 93.7 | 75.2 | 87.2 | 80.7 |
| G-net | 94.4 | 83.4 | 84.6 | **84.0** | 95.3 | 77.2 | 94.9 | **85.2** | 94.4 | 73.1 | 93.8 | 82.2 |
| R-net | 94.1 | 80.0 | 85.3 | 82.6 | 94.2 | 80.7 | 85.4 | **83.0** | 93.5 | 86.9 | 78.3 | 82.4 |

Table 8

Performance of Type 2 models in terms of Accuracy (A), Recall (R), Precision (P), and F1 score (F1) on the test dataset.

| Models | SP-V | | | | MS-V | | | | MFCC-V | | | |
|---|---|---|---|---|---|---|---|---|---|---|---|---|
| | A (%) | R (%) | P (%) | F1 (%) | A (%) | R (%) | P (%) | F1 (%) | A (%) | R (%) | P (%) | F1 (%) |
| V-net | 94.7 | 77.2 | 91.1 | 83.6 | 95.0 | 74.5 | 96.4 | 84.0 | 94.3 | 79.3 | 87.1 | 83.0 |
| G-net | 95.4 | 84.1 | 89.1 | **86.5** | 96.7 | 83.4 | 97.6 | **90.0** | 94.9 | 83.4 | 87.1 | 85.2 |
| R-net | 95.2 | 75.9 | 95.7 | 84.6 | 95.4 | 77.2 | 95.7 | **85.5** | 94.6 | 80.0 | 87.9 | 83.8 |

Table 9

Performance of Type 3 models in terms of Accuracy (A), Recall (R), Precision (P), and F1 score (F1) on the test dataset.

| Models | SP-V-A | | | | MS-V-A | | | | MFCC-V-A | | | |
|---|---|---|---|---|---|---|---|---|---|---|---|---|
| | A (%) | R (%) | P (%) | F1 (%) | A (%) | R (%) | P (%) | F1 (%) | A (%) | R (%) | P (%) | F1 (%) |
| V-net | 95.2 | 77.9 | 93.4 | 85.0 | 96.0 | 82.8 | 93.7 | 87.9 | 95.9 | 86.2 | 89.9 | 88.0 |
| G-net | 95.9 | 89.0 | 87.8 | 88.4 | 96.7 | 85.5 | 95.4 | **90.2** | 97.2 | 90.3 | 93.6 | **91.9** |
| R-net | 96.3 | 80.0 | 98.3 | 88.2 | 96.1 | 86.9 | 90.6 | **88.7** | 95.4 | 83.4 | 89.6 | 86.4 |

Table 10

Performance of Type 4 models in terms of Accuracy (A), Recall (R), Precision (P), and F1 score (F1) on the test dataset.

| Models | SP-MS | | | | SP-MFCC | | | | MS-MFCC | | | |
|---|---|---|---|---|---|---|---|---|---|---|---|---|
| | A (%) | R (%) | P (%) | F1 (%) | A (%) | R (%) | P (%) | F1 (%) | A (%) | R (%) | P (%) | F1 (%) |
| V-net | 95.2 | 81.4 | 90.1 | 85.5 | 94.8 | 79.3 | 89.8 | 84.2 | 95.0 | 78.6 | 91.9 | 84.8 |
| G-net | 96.4 | 92.4 | 87.6 | **89.9** | 96.0 | 82.1 | 94.4 | **87.8** | 95.5 | 77.9 | 95.8 | 85.9 |
| R-net | 96.1 | 81.4 | 95.9 | **88.1** | 95.2 | 85.5 | 86.7 | 86.1 | 95.7 | 76.6 | 98.2 | 86.0 |

Table 11

Performance of Type 5 models in terms of Accuracy (A), Recall (R), Precision (P), and F1 score (F1) on the test dataset.

| Models | SP-MS-MFCC | | | |
|---|---|---|---|---|
| | A (%) | R (%) | P (%) | F1 (%) |
| V-net | 96.1 | 92.4 | 86.5 | **89.3** |
| G-net | 96.3 | 85.5 | 92.5 | **88.9** |
| R-net | 96.5 | 91.0 | 89.2 | **90.1** |

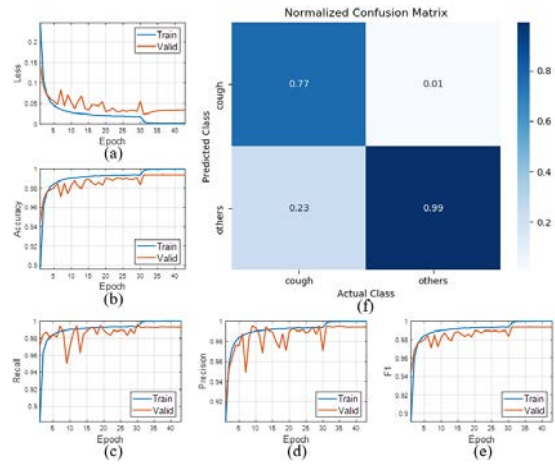

Fig. 9. Training results of the best Type 1 model (MS & G-net) with the F1 score of 85.2%: (a) Loss, (b) Accuracy, (c) Recall, (d) Precision, (e) F1 score, and (f) Normalized confusion matrix.

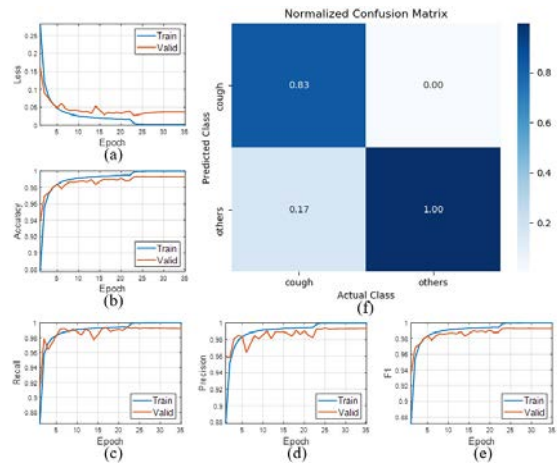

Fig. 10. Training results of the best Type 2 model (MS-V & G-net) with the F1 score of 90.0%: (a) Loss, (b) Accuracy, (c) Recall, (d) Precision, (e) F1 score, and (f) Normalized confusion matrix.

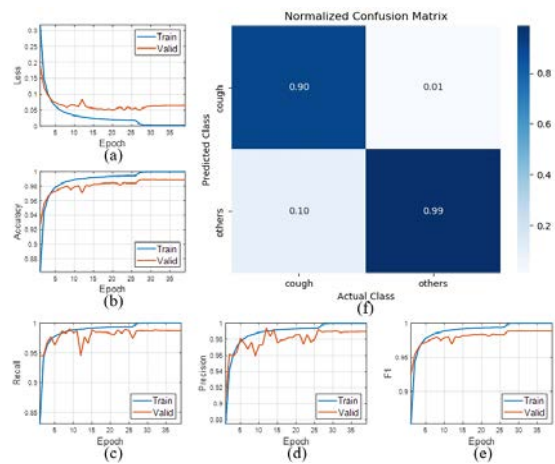

Fig. 11. Training results of the best Type 3 model (MFCC-V-A & G-net) with the F1 score of 91.9%: (a) Loss, (b) Accuracy, (c) Recall, (d) Precision, (e) F1 score, and (f) Normalized confusion matrix.

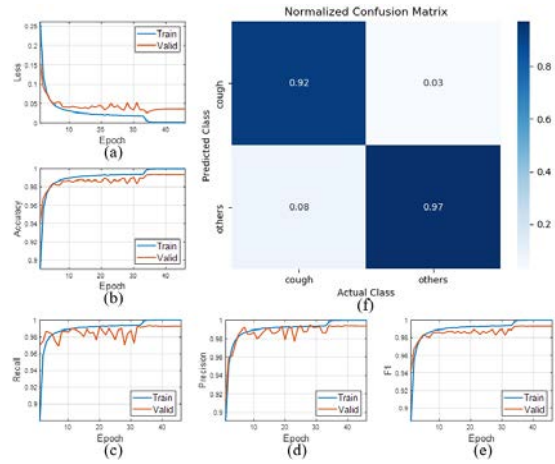

Fig. 12. Training results of the best Type 4 model (SP-MS & G-net) with the F1 score of 89.9%: (a) Loss, (b) Accuracy, (c) Recall, (d) Precision, (e) F1 score, and (f) Normalized confusion matrix.

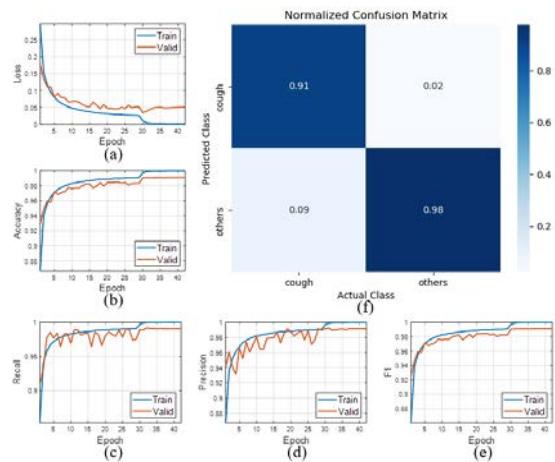

Fig. 13. Training results of the best Type 5 model (SP-MS-MFCC & R-net) with the F1 score of 90.1%: (a) Loss, (b) Accuracy, (c) Recall, (d) Precision, (e) F1 score, and (f) Normalized confusion matrix.

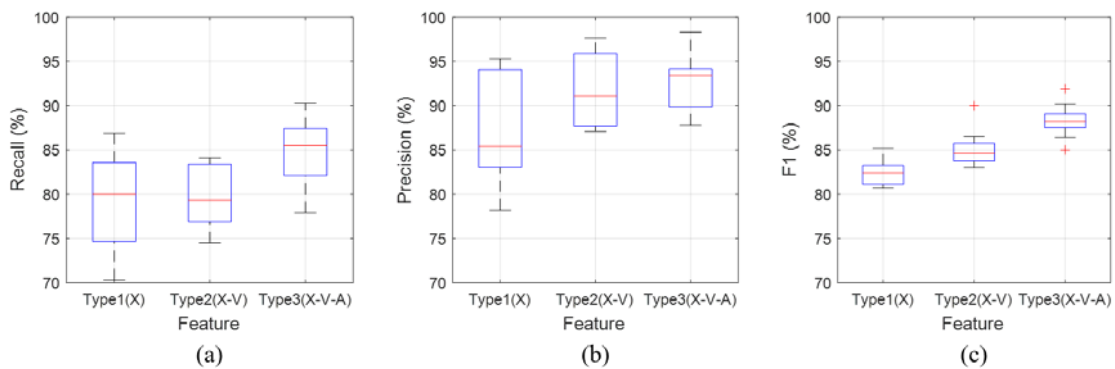

Fig. 14. Box-plots for (a) Recall, (b) Precision, and (c) F1 score showing performances on the test dataset of features in Types 1, 2, and 3 models. X indicates basic features of Type 1 including SP, MS, and MFCC. X-V denotes two-channel features of Type 2 in which X is concatenated by V-map (V). X-V-A represents three-channel features of Type 3 in which X is concatenated by V-map (V) and A-map (A).

but normalized $FN$ (0.09) slightly increased. Hence, SP-MS-MFCC & R-net shows performance similar to that of SP-MS & G-net.

When comparing each top model of Types 1, 2, and 3 shown in Tables 7, 8, and 9, MFCC-V-A & G-net of Type 3 showed the best performance with both recall and precision exceeding 90.0%. In addition, the top three results of Types 1, 2, and 3 showed that precision was higher than recall. High precision indicates low $FP$, which means that the problem of *Others* being misrecognized as *Cough* is alleviated. The box plots in Fig. 14 showed the changes in recall, precision, and F1 score according to the features of Types 1, 2, and 3 (in Tables 7, 8, and 9) to examine the effect of V-map and A-map on the performance. Type 3 features using both V-map and A-map showed the best performance in recall, precision, and F1 score. This indicates that V-map and A-map are effective in detecting the patterns of cough sounds that change rapidly in a short time. Overall, the precision was higher than the recall, suggesting that *Others* is less likely to be misrecognized as *Cough*. In addition, it was found that variances of recall and precision decreased as feature changed from Type 1 to Type 3. It means that using V-map and A-map to construct acoustic features of transient signals, such as cough sounds, shows less sensitive performance to basic features and networks. In cases of Types 4 and 5, it was found that these combinations of basic features were more effective than Type 1, and that the three-channel combination of Type 5 showed a slightly improved performance compared to the two-channel combination of Type 4. Overall, it was confirmed that using V-map and A-map (Type 3) is more effective in detecting coughing sounds than stacking three features (Type 5). Moreover, since FFT and related post-processing are required when calculating each basic feature, extracting the V-map and A-map from one basic feature is computationally more efficient than extracting three basic features individually.

Looking at the top model networks in Types 1 to 5, the 16-layer V-net was not selected once, whereas the 21-layer G-net was selected four times and the 50-layer R-net was selected once. Accordingly, it turns out that G-net captures the acoustic characteristics of cough sound much better than that of R-net, even though it used less than half of the layers compared to R-net. This is because G-net's inception module extracts various patterns simultaneously using kernels of different sizes, so it is effective in detecting large and small patterns of transient sounds such as coughing sounds.

Finally, MFCC-V-A & G-net were selected as the cough detection model, and it was integrated with a sound camera to localize cough sounds. Because MFCC-V-A & G-net had the highest F1 score and its recall and precision exceeded 90.0%, stable performance was expected in a subsequent pilot test.

**3. Sound camera**

The cough detection camera was functionally divided into a cough detection model that detects cough sounds and a sound camera that localizes sound sources. In this section, the principle of the sound camera is briefly introduced; then how it works with the cough detection model is explained.

3.1. Principle of the sound camera

A sound camera is an imaging device used to localize a sound source by using acoustic signals measured from a microphone array. Acoustic signals from the microphone array are processed using a beamforming method. The delay and sum (DAS) algorithm is one of the most common of the beamforming algorithms. The basic concept of the DAS beamforming method is shown in Fig. 15. When a sound from a source is measured by the microphone array, the output signals are usually aligned with specific time delays. If the delays in the signals are compensated exactly, they reinforce each other such that their summation can be maximized. In general, microphone arrays are highly directional, so sound sources can be localized by scanning in all possible directions. Therefore, the distribution of sound sources can be visualized by drawing their amplitude contours in the camera image. This is the main idea of the sound camera (Kim et al., 2014).

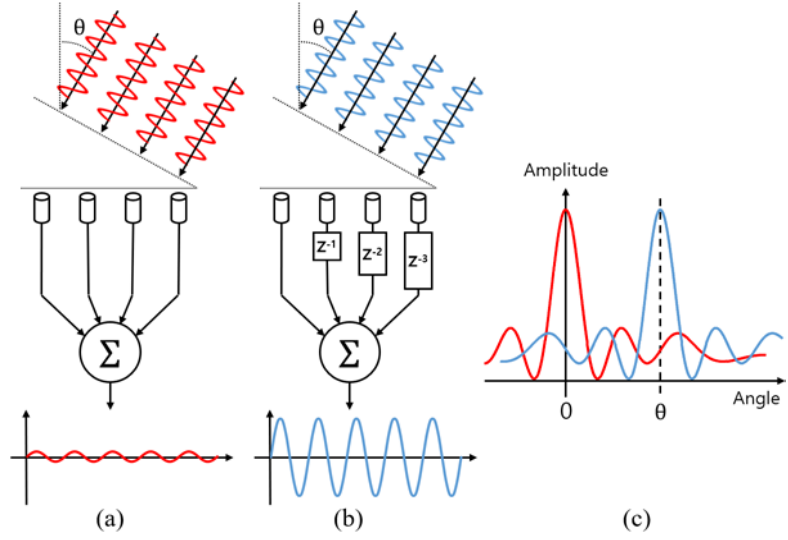

Fig. 15. Basic concept of delay and sum (DAS) beamforming method: (a) Cancellation of signals from an inclined direction, (b) Beam steering by DAS technique, and (c) Beamforming power distribution.

To explain the concept in more detail, Fig. 16 illustrates the DAS beamformer in the sound camera. In the time domain DAS processing, beamformer output can be defined as

$$B(t, \overrightarrow{X_P}) = \sum_{n=1}^{N} a_n s_n[t - \tau_n(\overrightarrow{X_P})] \tag{9}$$

where $t$ is time, $\overrightarrow{X_P}$ is the reference direction vector, $N$ is the total number of microphones, $a_n$ is a weight coefficient to be applied for individual microphones, $s_n(t)$ is the acoustic signal received by $n$-th microphone, and $\tau_n$ is a specified time delay to be imposed on the signal for virtual rotation of the array. In general, $a_n$ is set to unity, and $\tau_n$ is calculated through the following operation based on $\overrightarrow{X_P}$. First, the spatial domain of interest is discretized into pixels on a virtual inspection plane. Second, $\overrightarrow{X_P}$ is calculated for each pixel. Third, the position vector of the n-th microphone is obtained by $\overrightarrow{X_n} = \overrightarrow{X_P} - \overrightarrow{M_n}$, where $\overrightarrow{M_n}$ is the relative position vector of the $n$-th microphone from the origin of $\overrightarrow{X_P}$. Finally, the time delay of the $n$-th microphone, $\tau_n$ is obtained by

$$\tau_n(\overrightarrow{X_P}) = \frac{1}{c}(|\overrightarrow{X_P}| - |\overrightarrow{X_n}|) \tag{10}$$

where $c$ is the speed of sound (Rivey et al., 2017).

The scanning algorithm of the sound camera performs the delay calculation over the entire spatial domain it monitors. For a given pixel, after the time delays for all microphones are imposed on the input signals, the delayed signals are summed up as shown in Fig. 16(a) to produce a beamformer output. The amplitude contour can be obtained by representing the beamformer output amplitudes at all pixels as a color map. The contour is then overlaid on the captured camera image and output to the display on the sound camera. The location of sound sources is determined by the areas highlighted in the display. By taking several contoured images per second, the sound location information can be provided in video format. In this case, the location of a sound source can be visualized in a real time video.

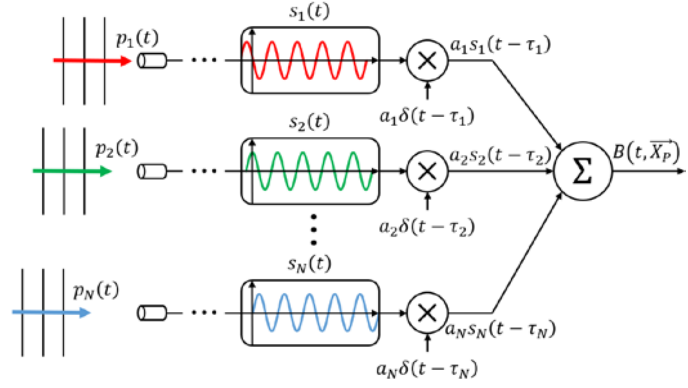

(a)

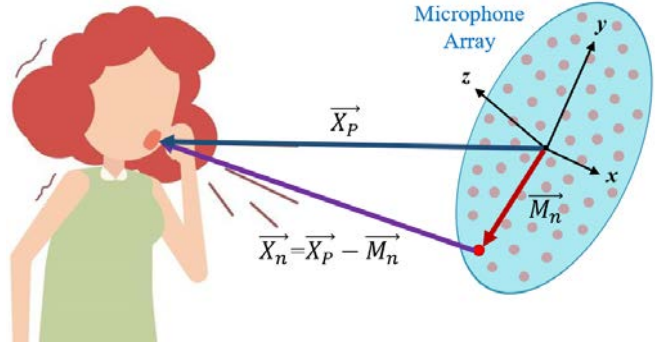

(b)

Fig. 16. Beamforming system of sound camera: (a) Structure of beamformer and (b) Location vectors for time delay calculation.

3.2. Sound camera integrated with cough detection model

A cough detection camera was developed by combining the cough detection model with BATCAM 2.0, a commercial sound camera developed by SM Instruments Inc. (BATCAM 2.0., 2021). As shown in Fig. 17, an optical camera (center) and an array of 112 microphones are mounted on the camera module as input devices on the front of the sound camera, and a touch screen added as an output device on the back. Because its sampling rate is 96 kHz, ultrasound up to 48 kHz can be measured. Detailed specifications of the sound camera are shown in Table 12.

The entire process of detecting coughing sounds and estimating their location with the cough detection camera is described in Fig. 18. The acoustic signals input to the microphone array are discretized through the data acquisition (DAQ) process, converted into an amplitude contour through the DAS beamforming process. Then, the sound camera waits for cough inference results from the cough detection model. While the sound is localized, the cough detection model extracts a feature (MFCC-V-A) from the input signal and outputs the cough inference result through its neural network (G-net). If the inference result is *Cough*, a contoured camera image with a "Cough" label is displayed on the screen, and if it is *Others*, only the camera image is displayed. In the case of cough in detail, one cough label with a red box is placed on the contour with the highest volume. The cough detection model receives a two-second buffered signal as an input. Since the input signal is formed by 75% overlapping, it is generated every 0.5 seconds. In addition, it takes about 0.025 seconds for the cough detection model to extract a feature and output an inference result. Since the input signal buffering and the cough detection process are performed in parallel, the cough detection camera outputs the inference result on the camera screen every 0.5 seconds. The total time may vary slightly depending on the buffering condition, and can be shortened by increasing the overlapping ratio. For example, when multiple people cough, the cough detection camera can detect their cough sounds if the cough sound onsets are more than 0.5 seconds apart from each other. Since a cough sound consists of a series of impulsive sounds, it is very difficult to match the onsets of several cough sounds in a short period of time. For this reason, the multiple coughs can be detected with reasonable time difference larger than 0.5 seconds. Therefore, if several people cough, the cough label in the camera image will appear alternately between coughing people with 0.5 second intervals.

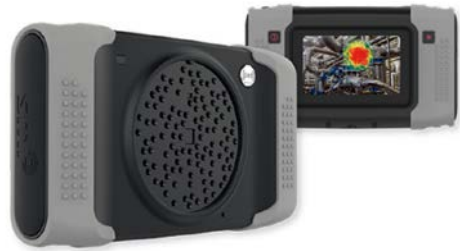

Fig. 17. The commercial sound camera (BATCAM 2.0) used to detect coughing sounds with an array of 112 MEMS microphones.

Table 12

Specifications of the sound camera used for cough detection.

| No. | Item | Value |
| --- | --- | --- |
| 1 | Microphone | 112 ch. digital MEMS |
| 2 | Effective frequency range | 2 k – 48 kHz |
| 3 | Microphone sensitivity | −41 dBFS |
| 4 | SNR | 66 dB(A) |
| 5 | Camera frame speed | 25 FPS |
| 6 | Detection distance | More than 0.3 m |
| 7 | Display type | 5" mono color LCD |
| 8 | Battery operation time | 4 hours |
| 9 | Product size | 237 mm × 146 mm × 56 mm |
| 10 | Product weight | 1.2 kg |
| 11 | Operating temperature | −20 – 50 ℃ |

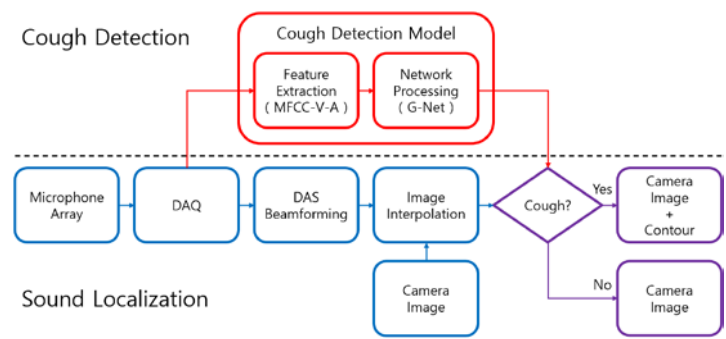

Fig. 18. Flowchart of cough detection and localization process.

Based on the above process, the proposed cough detection camera was developed, and a pilot test was conducted in a real office environment to verify its performance.

## 4. Results and discussion

This section presents the pilot test results of the proposed cough detection camera, and compares its performance with that of the previous prototype. Then, to analyze the cause of the performance, the pilot test results are discussed together with the training results of the best cough detection model.

4.1. Pilot test results

In a room with general office noise ranging from 45 to 55 dBA, the performance of the cough detection camera was tested by sequentially generating various sound events including coughing sounds. Table 13 shows the list of the sound events used in this pilot test. All the sound events were divided into two classes, "*Cough*" and "*Others*", consisting of 40 and 160 events, respectively. Moreover, "*Others*" had three subclasses: "cough-like sound", "impulsive sound", and "general sound". The first two are particularly important for the performance evaluation because they include sound events that can be mistaken for coughing sounds. Furthermore, the pilot test of the cough detection camera was conducted by comparing the performance of the proposed best model (MFCC-V-A & G-net) with that of a previous model (MS & C-net) (Lee et al., 2020). The previous cough detection model used MS as its input feature and output a cough inference result through C-net, a five-layer CNN architecture with GN. Reasonably, the 21-layer G-net is more beneficial than the five-layer C-

Table 13

List of sound events for pilot testing.

| Class | Subclass | Sound event type | Total |
|---|---|---|---|
| Cough | - | Cough | 40 |
| Others (10 per type) | Cough-like sound | Sneeze, throat-clearing | 20 |
| | Impulsive sound | Applause, clapping, knock, footsteps, door slam, finger snapping, chink & clink | 70 |
| | General sound | Speech, laughter, sniffle, snoring, page turn, phone, keyboard | 70 |

Table 14

Pilot test performance of cough detection camera in terms of Accuracy (A), Recall (R), Precision (P), and F1 score (F1).

| Model | A (%) | R (%) | P (%) | F1 (%) |
|---|---|---|---|---|
| Previous model (MS & C-net) | 84.0 | 90.0 | 56.3 | 69.2 |
| Proposed model (MFCC-V-A & G-net) | 96.0 | 90.0 | 90.0 | 90.0 |

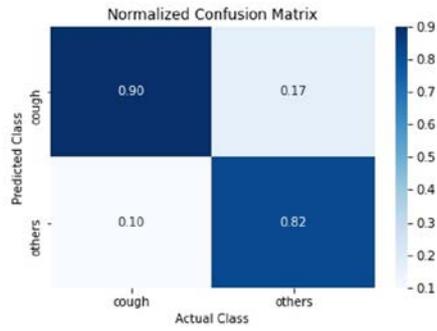

(a)

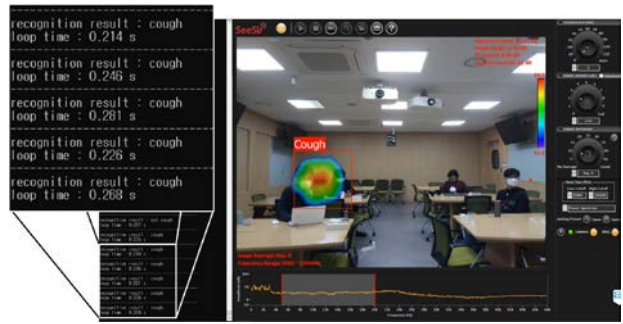

(b)

Fig. 19. Pilot test result of cough detection camera with previous model (MS & C-net): (a) Normalized confusion matrix and (b) Output screen of the cough detection camera.

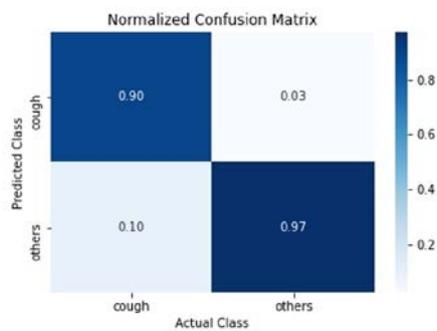

(a)

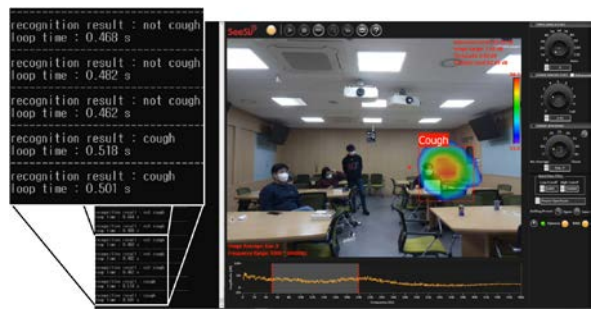

(b)

Fig. 20. Pilot test result of the cough detection camera with proposed model (MFCC-V-A & G-net): (a) Normalized confusion matrix and (b) Output screen of the cough detection camera.

net, but pilot tests were conducted on both models to verify the performance of the proposed model and to examine aspects where the previous model is still valid.

The pilot test results for the two models are shown in Figs. 19 and 20, respectively. In Figs. 19(b) and 20(b), the location and number of coughing sounds can be identified visually through the contour and label displayed on the output screen of the cough detection camera. In addition, a person who coughed could be tracked in real time by following the trajectory of the contour. From the confusion matrix results in Figs. 19(a) and 20(a), the normalized $TP$s of both models are equal to 0.9, indicating that 90% of the actual cough sounds were estimated correctly. However, in the case of normalized $TN$, the proposed model is 0.97, whereas the previous model is 0.82, which means that the proposed model 15% more accurate in predicting the class of sounds other than coughs. As briefly mentioned above, the previous model often misrecognized impulsive sounds as cough sounds, which reflects this result. In Table 14, both models exhibit 90.0% recall, indicating that the previous model is still valid in the ratio of accurately predicted cough sounds to actual cough sounds. However, in terms of precision, the previous model achieved 56.3%, while the proposed model achieved 90.0%, indicating that the proposed model is superior in its ratio of correctly predicted cough sounds to the total number of predicted cough sounds. Moreover, the proposed model achieved an F1 score of 90.0% (accuracy of 96.0%) in the pilot test. There was no significant difference in comparison with the F1 score of 91.9% (accuracy of 97.2%), which is the result for the test dataset in the modeling process. This reflects the robustness of the proposed model for different noisy environments because it was trained with augmented data from various background noises.

4.2. Discussion

The cough detection model was trained using datasets augmented with a background noise dataset (Dataset 7) so that the model would be less sensitive to ambient noise. As a result of the pilot test, the cough detection performance was found to be similar to that of the best model in the modeling process. In addition, cough detection suffers from data imbalance problem such that the number of cough data is generally much less than the number of non-cough data. The imbalanced data for each class can lead to convergence problem as it is difficult for the parameters of a neural network to converge. Therefore, to compensate for the data imbalance, cough data were generated with 50% overlap and then mixed with more background noise data considering the sizes of both cough and non-cough data in the training procedure. As a result, although the maximum number of epochs was set to 100, training of most models ended before 50 epochs.

As a result of the pilot test, it was found that the precision (90.0%) of the proposed model was significantly improved than that of the existing model (56.3%) as indicated in Table 14. The improved precision indicates a reduced $FP$, which means that the number of cases in which other sounds are misrecognized as cough sounds is decreased. The acoustic feature used in the proposed model was MFCC-V-A, which consists of MFCC and its V-map and A-map. MFCC has been frequently used in speech recognition because it mainly characterizes the spectral envelope of the vocal tract system. Since cough sound from the human mouth is filtered by the vocal tract, its spectral envelope can be accurately represented by the MFCC. Moreover, because V-map and A-map are the time derivatives of the frequency components in a spectrogram-based feature, the pattern of the frequency components that change over time is clearly revealed. Therefore, MFCC-V-A is very effective in detecting the acoustic patterns of cough sounds. In an auditory cortex model, auditory features are extracted in a transition process from acoustic features to perceptual features and finally to category-specific representations at the top level (Bizley & Cohen, 2013). This is similar to the biological processing of vision information. Moreover, human brains tend to respond sensitively to the rate of change. Reflecting the human auditory perception process, the three-channel spectrogram-based feature with V-map and A-map was named *Spectroflow*.

The network applied to the cough detection camera was G-net, consisting of nine inception modules. In each inception module, kernels of various sizes (e.g., $1 \times 1$, $3 \times 3$, and $5 \times 5$) extract feature maps simultaneously. Since cough sound is transient, its energy decreases over time, and the low-frequency components last relatively longer than the high-frequency components. These various decay times appear as different lengths of pattern on the spectrogram-based acoustic features. It would be effective to use a combination kernel of various sizes to extract acoustic patterns of different sizes simultaneously. Therefore, it is thought that the inception module of G-net was effective in detecting the pattern change of cough sound from various sounds.

5. Conclusion

Since the outbreak of COVID-19, there has been increasing demand for systems that can detect infection symptoms in real time in the field. It is very important to detect coughs to prevent the spread of infectious diseases, because droplets released during coughing are one of the transmission pathways of viral disease. From this perspective, a cough detection camera was developed that can monitor coughing sounds in real time in the field, and its detection performance was evaluated by conducting a pilot test in an office environment. As a result of reviewing the modeling process and the pilot test, we confirmed that DA technique, *Spectroflow*, and the inception module of G-net have significant contributions to the performance of the cough detection camera. For future works, it is suggested to improve further the cough detection performance by reflecting real environmental noise or proposing a network that maximizes the advantages of the inception module. In addition, through data collection using IoT devices, the cough detection camera is expected to be used as a medical device that automatically monitors patient conditions in hospitals or as a monitoring system to detect epidemics in public places, such as schools, offices, and restaurants.

**Acknowledgment**

This work was supported by the "Human Resources Program in Energy Technology" of the Korea Institute of Energy Technology Evaluation and Planning (KETEP), granted financial resources from the Ministry of Trade, Industry & Energy, Republic of Korea (No. 20204030200050), and also supported by a Korea Institute of Marine Science and Technology Promotion (KIMST) grant funded by the year 2022 Finances of the Korea Ministry of Oceans and Fisheries (MOF) (Development of Technology for Localization of Core Equipment in the Marine Fisheries Industry, 20210623).